\documentclass[preprint,amsmath,amssymb]{revtex4}
\usepackage{latexsym}
\usepackage[dvips]{graphicx}
\usepackage{bm}

\begin{document}

\title{Higgs boson decays in the littlest Higss model}
\author{G. A. Gonz\'alez-Sprinberg (1), R. Mart\'\i nez (2) and  J.-Alexis 
Rodr\'\i guez (2)}

\altaffiliation{gabrielg@fisica.edu.uy, 
remartinezm@unal.edu.co, jarodriguezl@unal.edu.co}
\affiliation{(1) Instituto de F\'{\i}sica,
 Facultad de Ciencias, Universidad de la Rep\'ublica,  Igu\'a 4225,\\ 
11400 Montevideo, Uruguay\\ \\
(2) Departamento de F\'\i sica, Universidad Nacional de Colombia \\ 
Bogot\'a, Colombia}

\date{}

                                                                             
\begin{abstract}
We calculate the two body Higgs boson decays in the framework of 
the littlest Higgs model. The decay $H \to \gamma Z$ is computed at one 
loop level and,  using previous results, we evaluate the branching 
fractions 
in the framework of the littlest Higgs
 model. A wide range of the space parameter of the model is 
considered
 and possible deviations from the standard model are explored.
\end{abstract}

\maketitle

Although the standard model (SM) has passed tests up to the 
highest energies accessible today, it remains unanswered the 
mechanism responsible for the generation of mass and 
behind the spontaneous breaking of the $SU(2)_L \times U(1)_Y$ symmetry. 
The precision electroweak measurements indicate a possible light  Higgs boson \cite{lep},  
but the SM suffers of the so-called hierarchy or fine-tuning problem which is 
due to the presence of quadratic divergences in the loop processes for the scalar
 Higgs boson self-energy. This lead to the open question of how to prevent the Higgs 
 boson mass from getting grand unification theory (GUT) scale contributions, in order to
  remain it light. An answer has come from supersymmetry (SUSY), where the introduction 
  of supersymmetric partners with opposite statistics of the existing particle cancel out 
  the divergences \cite{susy}. Strong dynamics at the TeV-scale has also been invoked by 
  theories, such as  topcolor models \cite{topcolor} where the electroweak symmetry breaking 
  is done dynamically. Recently has appeared a new proposal, 
  called Little Higgs models, which could solve the problem in a natural way by making the 
  Higgs boson a pseudo-Goldstone boson, whose mass is protected to get quadratic 
  divergences from the gauge sector by a global symmetry \cite{lh}.

In the following we use the littlest Higgs model  which is based on a global
 $SU(5)$ symmetry \cite{lh2}. It is a non-linear sigma model which is broken 
 when the symmetric tensor gets an expectation value, down the symmetry to 
 global $SO(5)$. At the same time, the subgroup $[SU(2) \times U(1)]^2$ of $SU(5)$ is gauged 
 and broken to its diagonal subgroup $SU(2)_L \times U(1)_Y$, the SM symmetry.
  The key point is in the restoration of part of the global symmetry by ungauging
   some part of the gauge symmetry. In this model the restored symmetry is $SU(3)$.
    Therefore the Higgs boson remains light because it becomes a Goldstone boson of 
    spontaneously broken $SU(3)$ and acquires a mass at the electroweak scale by 
    radiative corrections. Hence, the Higgs boson mass does not acquire contributions 
    from gauge loops because it is protected by an approximate global symmetry and 
    the quadratic divergences are cancel off because the gauge sector of the two
     $SU(2) \times U(1)$ are coupling with the Higgs boson with opposite sign.
 
On the other hand, the contributions to the Higgs boson mass from interactions with fermions and from 
quartic couplings, should obey the approximate global symmetries. Thus, to control the 
quadratic divergences coming from the top quark coupling, the model  adds a new 
like-vector  T-quark and then allows  the top quark to get a mass from a collective
 breaking. These cancellations allow that the little Higgs theory be valid up to a
  scale of the order of 10 TeV without any fine-tuning. At this scale the theory 
  becomes a strongly interacting one \cite{lh, lh2}.
At the end, in the littlest Higgs model there are four new gauge bosons $W_H^a$ and $B_H$,
 a vector-like T-quark and a new triplet scalar field of $SU(2)_L$, all of them playing 
 together in order to successfully cancel off the quadratic divergences of the Higgs boson 
 self-energy\cite{lh, lh2, logan1}.

The effects of the new states at low energies have been studied and 
several constraints on the model parameters have been imposed in the literature. A very exhaustive 
study of this model was presented in reference \cite{logan1}.  Different studies using
 precision electroweak measurements \cite{lh3} and possible signatures at future 
 colliders have also been done \cite{lhc-lh}.

One important job for the present and future experiments is to search for the Higgs 
boson and investigate its properties. But, even in the framework of the SM the Higgs 
boson has several possibilities to be produced. Depending on the Higgs boson mass,
 different decay modes open up \cite{hhg}. In the framework of the SM for instance,
  the decay to $b \bar b$ is dominant but the decay $h \to \tau^+ \tau^-$ is also 
  sizable. With increasing mass, the Higgs boson  decays heavier particles become dominant, 
  mostly into the channels with $ZZ$  and $W^+ W^-$ pairs. However, the ratio of $h\to \gamma \gamma$
   decay, although it is small one because it is at one-loop level, can be important 
   below the $W^+ W^-$ pair threshold because of its well-defined final state. Thus, the
    production processes at colliders in combination with all opened decay channels 
    result in a large number of possible final states. Our aim is to study the Higgs boson
     decays in the framework of the littlest Higgs model using as a paradigm the SM framework.
      In this way we can compare the different modes and we can use different constraints on the 
      parameter space taken from the literature in the framework of the littlest Higgs model.

The littlest Higgs (LH) model is based on an $SU(5)/SO(5)$
non-linear sigma model. A vacuum expectation value (VEV) breaks the 
$SU(5)$ global symmetry into its subgroup
$SO(5)$ and breaks the local gauge symmetry $[SU(2) \otimes
U(1)]^2$ into its diagonal subgroup $SU(2)_L \otimes U(1)_Y$ at
the same time, which is identified as the SM electroweak gauge
group. At the scale $\Lambda_s \sim 4 \pi f$, the VEV
 associated with the spontaneous
symmetry breaking, proportional to the scale $f$, is parameterized
by 
\begin{equation}
\Sigma_0 = \left( \begin{array}{ccc}
 & & {\bf{1}}_{2 \times 2} \\
 &1 & \\
{\bf{1}}_{2 \times 2} & &
\end{array}\right) .
\end{equation}
 The scalar fields  of the non-linear $\sigma$-model can be written as
\begin{eqnarray}
\Sigma(x) = e^{i \Pi(x)/f} \Sigma_0 e^{i \Pi(x)^{T}/f},
\end{eqnarray}
where $\Pi(x) = \pi^a(x) X^a$ is the Goldstone boson matrix. $X^a$
are the broken generators of $SU(5)$.  The Goldstone boson matrix $\Pi(x)$ can be expressed as
\begin{equation}
\Pi=\left( \begin{array}{ccc} {}& h^\dagger/\sqrt 2&
\phi^\dagger\\h/\sqrt 2 &{}& h^*/\sqrt 2\\\phi  & h^T/\sqrt 2 &{}
    \end{array}\right),
\end{equation}
where
\begin{eqnarray}
h = \left(
    \begin{array}{cc}
    h^{+} ~~ h^0
    \end{array}
    \right), ~~~~~~
\phi = \left(
       \begin{array}{cc}
       \phi^{++} & \phi^{+}/\sqrt{2} \\
       \phi^{+}/\sqrt{2} & \phi^0
       \end{array}
       \right)
\end{eqnarray}
are doublet and triplet under the unbroken SM gauge group $SU(2)_L \otimes
U(1)_Y$, respectively.

The leading order dimension-two term for the scalar fields
$\Sigma(x)$ in the littlest Higgs model can be written as
\begin{equation}
{\cal L} = \frac{1}{2} \frac{f^2}{4} {\rm Tr} |{\cal D}_{\mu}
\Sigma |^2,
\end{equation}
where ${\cal D}_{\mu}$ is the covariant derivative for the gauge group
$[SU(2) \otimes U(1)]^2 = [SU(2)_1 \otimes U(1)_1] \otimes
[SU(2)_2 \otimes U(1)_2]$ and can be written in the following way
\begin{equation}
{\cal D}_\mu \Sigma =
\partial_\mu \Sigma - i \sum_{j=1}^2
\left[ g_j ( W_j \Sigma + \Sigma W_j^T) + g_j^{\prime} (B_j \Sigma
+ \Sigma B_j^T) \right],
\end{equation}
where $W_{\mu j} = \sum_{a = 1}^3 W_{\mu j}^a Q_j^a$ and $B_j =
B_{\mu j} Y_j$ are the $SU(2)_j$ and $U(1)_j$ gauge fields,
respectively. The generators of two $SU(2)$'s ($Q_j^a$) and two
$U(1)$'s generators ($Y_j$) are as follows
\begin{eqnarray}
Q_1^a = \left(
        \begin{array}{cc}
        \frac{\sigma^a}{2} & \\
        & {\bf{0}}_{3 \times 3}
        \end{array}
    \right),& ~~&
Q_2^a = \left(
        \begin{array}{cc}
        {\bf{0}}_{3 \times 3} & \\
        & -\frac{\sigma^{a \ast}}{2}
\end{array}\right),  \\
Y_1 = {\rm diag}\{-3,~ -3,~ 2,~ 2,~ 2\}/10, &~~& Y_2 = {\rm
diag}\{-2,~ -2,~ -2,~ 3,~ 3\}/10,\nonumber
\end{eqnarray}
where $\sigma^a~ (a = 1, 2, 3)$ are the Pauli matrices. As we
expect, the breaking of the gauge symmetry $[SU(2) \times U(1)]^2$
into its diagonal subgroup $SU(2)_L \times U(1)_Y$ gives rise to
heavy gauge bosons $W^{\prime}$ and $B^{\prime}$, and the 
unbroken subgroup $SU(2)_L \times U(1)_Y$ introduces the massless
gauge bosons $W$ and $B$.

In the littlest Higgs model there is no Higgs
potential at tree-level. Instead, the Higgs potential is generated
at one-loop level and higher orders due to the interactions with gauge
bosons and fermions.  The Higgs
potential  can be presented in the standard form of a Coleman-Weinberg potential as
\begin{eqnarray}
\label{win} V &=& \lambda_{\phi^2} f^2 {\rm Tr}( \phi^{\dag} \phi)
+ i \lambda_{h \phi h} f (h \phi^{\dag} h^T - h^{\ast} \phi
h^{\dag}) - \mu^2 h h^{\dag} +
\lambda_{h^4} (h h^{\dag})^2 \\
& & + \lambda_{h \phi \phi h} h \phi^{\dag} \phi h^{\dag} +
\lambda_{h^2 \phi^2} h h^{\dag} {\rm Tr}(\phi^{\dag} \phi) +
\lambda_{\phi^2 \phi^2} \left(
                        {\rm Tr}(\phi^{\dag} \phi)
            \right)^2 +
\lambda_{\phi^4} {\rm Tr}(\phi^{\dag} \phi \phi^{\dag} \phi),\nonumber 
\end{eqnarray}
where $\lambda_{\phi^2}$, $\lambda_{h \phi h}$, $\lambda_{h^4}$,
$\lambda_{h \phi \phi h}$, $\lambda_{h^2 \phi^2}$,
$\lambda_{\phi^2 \phi^2}$ and $\lambda_{\phi^4}$ are the
coefficients of the original Higgs potential.
By minimizing the Coleman-Weinberg potential, we obtain the vacuum
expectation values $\langle h^0 \rangle =v/\sqrt 2$, $\langle i
\phi^0 \rangle=v^{\prime}$ of the Higgs boson doublet $h$ and
triplet $\phi$, which give rise to the electroweak symmetry
breaking (EWSB). In order to get the correct vacuum for the  
electroweak symmetry breaking with $m_H^2>0$, it is possible to 
express all four parameters in the Higgs potential to leading order 
in terms of the physical parameters $f$, $m_H^2$, $\nu$ and $\nu'$. After EWSB, the gauge
sector gets additional mass and mixing term due to the VEVs of $h$
and $\phi$. By diagonalizing the quadratic term of the gauge
sector, we may get the mass eigenstates for the light bosons $A_L$, $Z_L$, $W_L$,
and for the heavy ones, $A_H$, $Z_H$ and $W_H$, with masses of the order 
of $\nu^2 /f^2$. To leading order they are
\begin{eqnarray}
M_{W_L}^2 &=& m_w^2 \left [ 1- \frac{\nu^2}{f^2} \left 
(\frac 1 6+ \frac 1 4 (c^2-s^2)^2- \frac{x^2} {4}  \right )   \right ] , \nonumber\\
M_{W_H}^2 &=& m_w^2 \frac {f^2}{s^2 c^2 \nu^2} ,\nonumber\\
M_\Phi^2 &=& \frac{2 m_H^2 f^2}{(1-x^2) \nu^2},
\end{eqnarray}
where $m_w=g \nu /2$; the mixing between the two gauge groups $SU(2)_i$ is 
parametrized by $c$ and in  the Higgs sector the parameter $x=4f \nu'/v^2$ is defined.
\par
 Large quadratic divergence in the Higgs boson mass due to
the heavy top Yukawa interaction is a problem in the SM. The littlest Higgs model 
 solve this problem by introducing a pair of new
fermions $\tilde{t}$ and $\tilde{t}^{\prime}$ 
which are a vector-like pair. Their interactions are included in the following Lagrangian:
\begin{equation}
\label{L1}{\cal L}_Y = \frac{1}{2}\lambda_1 f
\epsilon_{ijk}\epsilon_{xy}\chi_i \Sigma_{jx}\Sigma_{ky}u^{\prime
c}_3+ \lambda_2f \tilde{t}\tilde{t}^{\prime c}+h.c.
\end{equation}
where $\chi=(b_3,~ t_3,~\tilde{t})$, $\epsilon_{ijk}$ and
$\epsilon_{xy}$ are antisymmetric tensors where $i$, $j$, $k$ run
through 1, 2, 3 and $x$, $y$ run through 4, 5, and $\lambda_1$ and
$\lambda_2$ are the new model parameters. By expanding the
Lagrangian (\ref{L1}) and diagonalizing the mass matrix, we get the physical states of the top quark $t$ and a
new heavy vector-like quark $T$. The masses of the two physical
states are
\begin{eqnarray}
\label{mt1} m_t = \frac{ v \lambda_1\lambda_2 }{\sqrt{\lambda_1^2
  + \lambda_2^2}} \left\{ 1+\frac{v^2}{f^2} \left [-\frac{1}{3}+\frac{fv'}{v^2}+
  \frac{1}{2}\frac{\lambda_1^2}{\lambda_1^2+\lambda_2^2} \left(
  1-\frac{\lambda_1^2}{\lambda_1^2+\lambda_2^2} \right ) \right ] \right\} ~ ,
\end{eqnarray}
\begin{eqnarray}
\label{mt2}m_T = \frac {m_t f}{s_t c_t \nu} ,
\end{eqnarray}
respectively, and the mixing between the top quark and the heavy vector-like 
quark $T$ is parametrized by $c_t$. Since the top quark mass is already obtained in the
SM, we can then get the parameter relation from Eq.(\ref{mt1}) 
\begin{eqnarray}
\label{l1l2}\frac{1}{\lambda_1^2}+\frac{1}{\lambda_2^2}\approx
\frac{v^2}{{m_t}^2} \approx 2.
\end{eqnarray}

In general, all the couplings to the Higgs boson in the Littlest
 Higgs model are modified at order $\nu^2/f^2$, so we use the following parameterization,
\begin{eqnarray}
{\cal L}&=&- \frac {m_t}{\nu} y_t \bar t t H -\frac {m_t}{\nu} y_T \bar T T H + 
2 \frac {m_{W_{L}}^2}{\nu} y_{w_L} W_L^+ W_L^- H + 2 \frac {m_{W_{L}}^2}{\nu} y_{w_H} W_H^+ W_H^- H \nonumber \\
&-& 2 \frac {m_{\Phi^+}^2}{\nu} y_{\Phi^+} \Phi^+ \Phi^- H -\frac{m_b}{\nu} 
y_b \bar b b H + \frac {m_{Z_L}^2}{\nu} y_{Z_L} Z_L Z_L H
\end{eqnarray}
where the $y_i$ factors are derived from the Higgs couplings given in ref 
\cite{logan1} and they can be written as
\begin{eqnarray}
y_t &=& 1 + \frac {\nu^2}{f^2}(- \frac 2 3 +\frac 1 2 x - \frac 1 4 x^2 +c_t^2(1+c_t^2))\nonumber \\
y_b &=& 1 + \frac {\nu^2}{f^2}(- \frac 2 3 +\frac 1 2 x - \frac 1 4 x^2) \nonumber \\
y_T &=& \frac {\nu} f c_t (1+c_t^2) \nonumber \\
y_{w_L}&=& 1+\frac {\nu^2}{f^2}(-\frac 1 6 + \frac 3 4 (c^2-s^2)^2-x^2 ) \nonumber \\
y_{Z_L}&=& 1+\frac {\nu^2}{f^2}(-\frac 1 6 - \frac 1 4 (c^2-s^2)^2-\frac 5 4 (c'^2-s'^2)^2+ \frac 1 4 x^2 )
\end{eqnarray}
The partial widths will be proportional to the above factors $y_i$ or combinations of them. 


The width of the tree level decay $H \to \bar f f$ is
\begin{equation}
\Gamma(H \to ff)^{LH}= \frac {N_c}{8 \pi} \sqrt{2} G_F m_f^2 \beta^3 m_H y_f^2 y_{G_F}^2
\end{equation}
where
\begin{equation}
y^2_{G_F} = 1 + \frac{\nu^2}{f^2} (-\frac 5 {12} + \frac 1 4 x^2)
\end{equation}
We use  only the $H \to \bar b b$ decay coming from LH model, while the other options
 like the decay into leptons are taken as in the SM model.
  Here we should note that the factors $y_i$ for the standard particles 
   go to one when new physics is turn off, but they become zero for the new 
   particles of the model. This means that the SM expressions can be obtained 
   asymptotically when the new physics is decoupled in the limit $f \to \infty$. The decays into WW or ZZ bosons 
   are like the SM ones, but they get a new factor $y_{WW}$ or $y_{ZZ}$ respectively,\\
\begin{eqnarray}
y_{WW} &=&  1 + \frac {\nu^2}{4 f^2}(-3 +6 (c^2 -s^2)-7 x^2)\nonumber \\
y_{ZZ} &=& 1 + \frac {\nu^2}{4 f^2}(-3 -2(c^2 -s^2)-10(c'^2-s'^2) + 3 x^2)
\end{eqnarray}
These factors are coming from products of $y_i$ related with the corrections of new physics. 
We use the W  boson mass as an input, not the Z boson mass; note that 
the factor $y_{WW}$ is different of the factor used in reference \cite{logan3}.

The one loop-level  decays $H \to \gamma \gamma$ and $H \to g g$ are taken from reference \cite{logan1} and the decay $H \to \gamma Z$ is
\begin{equation}
\Gamma(H \to \gamma Z) = \frac{\sqrt{2} G_F \alpha^2 m_h^3 y^2_{G_F}}{128 \pi^3} (1- \frac{m_Z^2}{m_h^2})^3 \vert A_F + A_W \vert^2
\end{equation}
where

\begin{eqnarray}
A_F &=& \frac {-2}{s_w c_w} y_{top} (I_1(\tau_t,\lambda_t) -I_2(\tau_t,\lambda_t)) - \frac 2 3  t_w y_T (I_1(\tau_T,\lambda_T)
-I_2(\tau_T,\lambda_T))  \nonumber \\
A_W &=& \cot_w \left \{ y_{W_L} \left[ I_1(\tau_w,\lambda_w)( -4+2 t_w^2- \frac {m_h^2}{m_{W_L}^2} \frac{c_{2w}}{2 c_w^2}
-\frac 1{c_w^2})\right. \right.\\
&+& \left. 4 I_2(\tau_w,\lambda_w )(4- \frac 1 {c_w^2})   \right ]   \nonumber \\
&+&  y_{W_H}  \left [ I_1(\tau_{W_H},\lambda_{W_H})( -4+2 t_w^2 (1+ \frac{m_{W_L}}{m_{W_H}}) \frac{m_{W_L}}{m_{W_H}}
 - \frac {m_h^2}{m_{W_H}^2} \frac{c_{2w}}{2 c_w^2})\right. \nonumber \\
&+& \left.\left. 4 I_2(\tau_{W_H},\lambda_{W_H})(4- \frac {2 m^2_{W_L} t_w^2}{m^2_{W_H}}) \right ] +   y_{\phi^+} t_w^2 \left [ 2 I_2 (\tau_{\phi^+},\lambda_{\phi^+})-I_1(\tau_{\phi^+},\lambda_{\phi^+}) \right ] \right\} \nonumber 
\end{eqnarray}
the integrals $I_1$ and $I_2$ can be found in \cite{hhg}.  The arguments depen on 
 the masses where the integrals are to be evaluated, and the new factor
\begin{eqnarray}
y_{top}&=&\frac 12 - \frac 43 s_w^2 + \frac{\nu^2}{f^2}(-\frac 2 3+c_t^2(2+c_t^2)+ \frac 12 x 
-\frac 14 x^2 +\frac 12 c^2 (c^2-s^2)\nonumber \\
&-&\frac 5 2 (c'^2-s'^2)(-\frac 4 5+c'^2 -c_t^2(\frac {20} {15} -\frac 23 c'^2)))
\end{eqnarray}
is the result of the product of $y_t$ and the correction of the $\bar t t Z_L$ coupling.

The results are shown in figures 1-5. The regions between lines correspond to accessible values
 on some parameters. The parameters should be taken in the intervals
$0 \leq c_t \leq 1$, $0 \leq  x \leq 1$ and $0.4 \leq c' \leq 1$. Figure 1 shows the ratio
 $BR(H \to \bar b b )^{LH}/BR(H \to \bar b b )^{SM}$ versus the scale $f$; this ratio
  depends on different parameters including the Higgs boson mass because it is the 
  ratio between the branching fraction. For $f=1$ TeV, $m_H=120$ GeV, the branching 
  fraction of the littlest Higgs model is modified from $+2\%$ to $-8\%$.
   As we expect, the region is converging to one as $f$ increase, where the new physics is decoupling.
    If we consider the ratio between the widths of $H \to \bar b b$, the Higgs
     boson mass dependence cancels out, as shown in figure 3. 
     Figure 2 presents the dependence of the ratio between the branching fractions with 
     the Higgs boson mass,  where we have fixed the scale at $f=1$ TeV. In that plot we note 
     that for $m_h= 180$ GeV the branching can be deviated $\pm 9\%$ from the SM value.
      Figure 3 shows the case of ratio of  widths for $H \to \bar b b$ and, as 
      already mentioned, it does not depend on the Higgs mass. For $f=1$ TeV the decay
       in the framework of the littlest Higgs model is suppressed respect to the SM 
       between $11\%$ and $6 \%$; this is because of the factor $y_b^2 y_{G_F}^2$, that  has
        a negative term of the order of $\nu^2/f^2$. When $f$ grows up the rate is 
	getting closer to one, as expected.

Figure 4 shows the ratio between the widths for the one-loop level decay $H \to \gamma Z$
 against the scale $f$; we expect a deviation about $+ 9 \%$-$-15 \%$ for $f=1$ TeV. Finally, 
 figure 5 shows the branching fractions for a wide range of parameters, fixed  $f=1$ TeV versus
  the Higgs boson mass. The curves are showing zones where the parameters of the model are valid.
   In fact the branching fraction of the SM would be inside the region.

 In conclusion, we estimate the branching fraction of the different channels for the Higgs 
 boson  in the framework of the littlest Higgs model in particular we show the one-loop level 
 $H \to \gamma Z$ and in different figures we show the behavior respect to the SM and in
  general deviations about 10\% are expected.

{\it Note added: During the elaboration of this work another paper 
\cite{logan3} on a similar subject appeared}

We acknowledge the financial support from COLCIENCIAS.

\newpage

\begin{figure}
\includegraphics[scale=0.5, angle=-90]{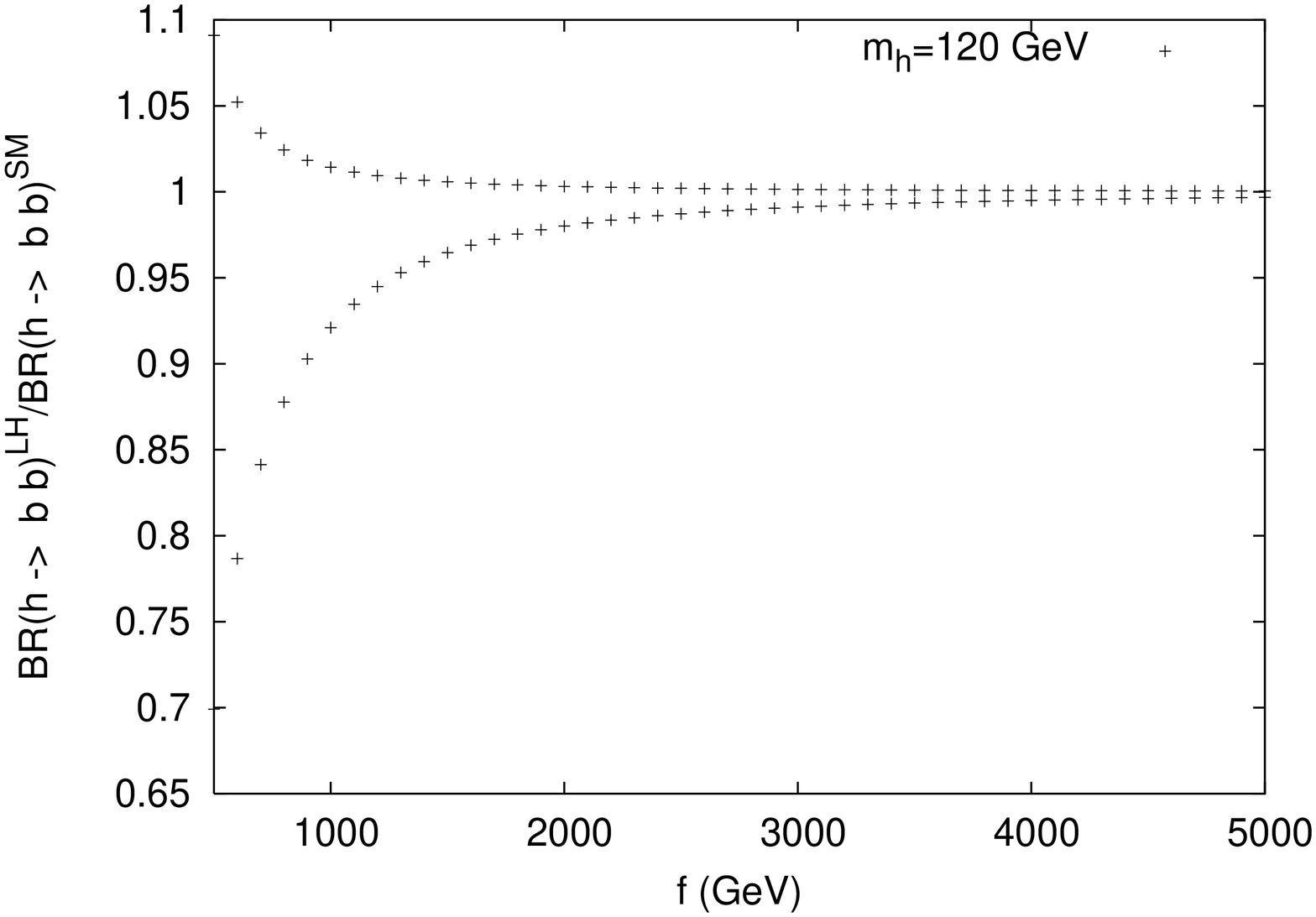}
\caption{The ratio
 $BR(H \to \bar b b )^{LH}/BR(H \to \bar b b )^{SM}$ versus the scale $f$, $m_H=120$}
\end{figure}

\begin{figure}
\includegraphics[scale=0.5, angle=-90]{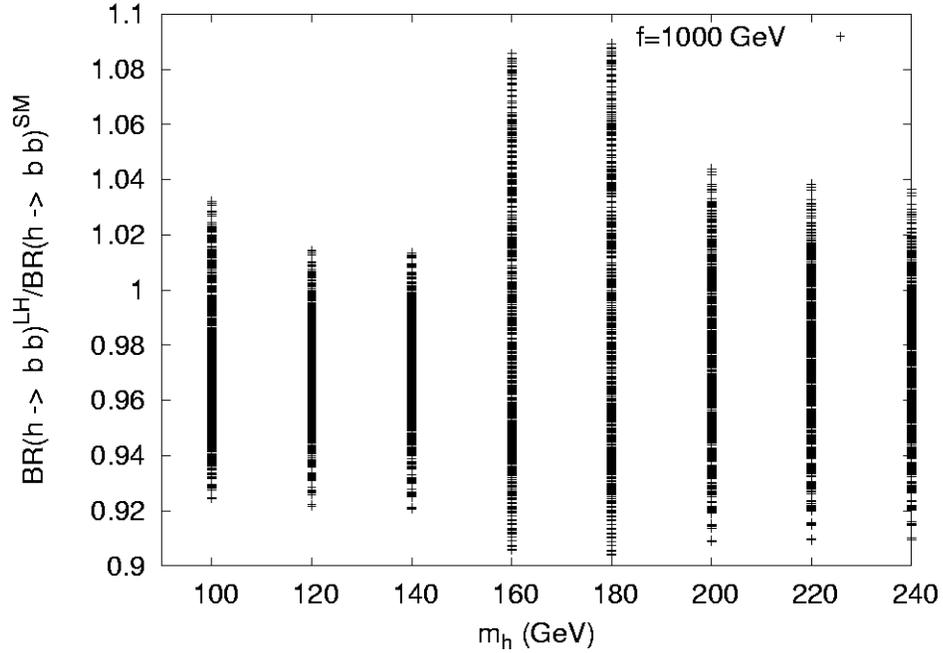}
\caption{ The ratio between the branching fractions $BR(H \to \bar b b )^{LH}/BR(H \to \bar b b )^{SM}$ versus
     the Higgs boson mass,  where we have fixed the scale at $f=1$ TeV. }
\end{figure}
                                            
\begin{figure}
\includegraphics[scale=0.5,angle=-90]{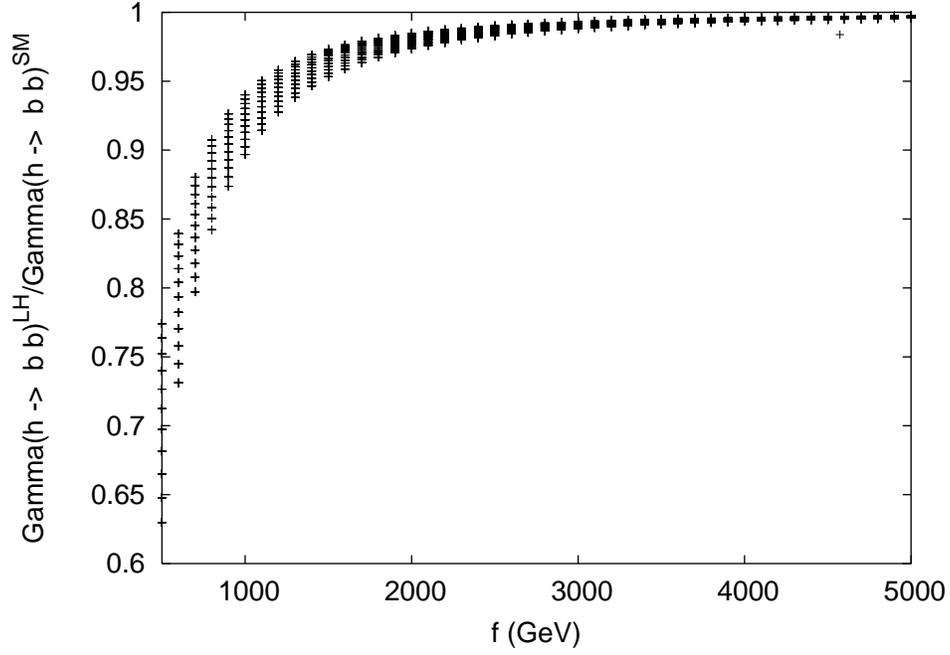}
\caption{The ratio of  widths for $H \to \bar b b$ versus the scale $f$, it does not depend on the Higgs mass. }
\end{figure}

\begin{figure}
\includegraphics[scale=0.5,angle=-90]{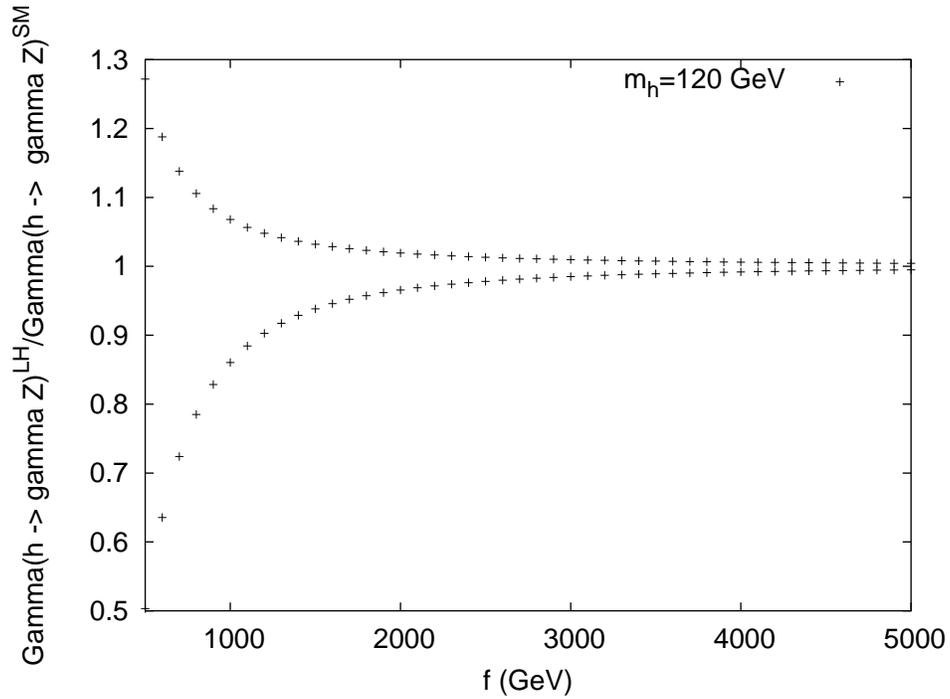}
\caption{The ratio between the widths for the one-loop level decay $H \to \gamma Z$
 against the scale $f$. }
\end{figure}

\begin{figure}
\includegraphics[scale=0.5,angle=-90]{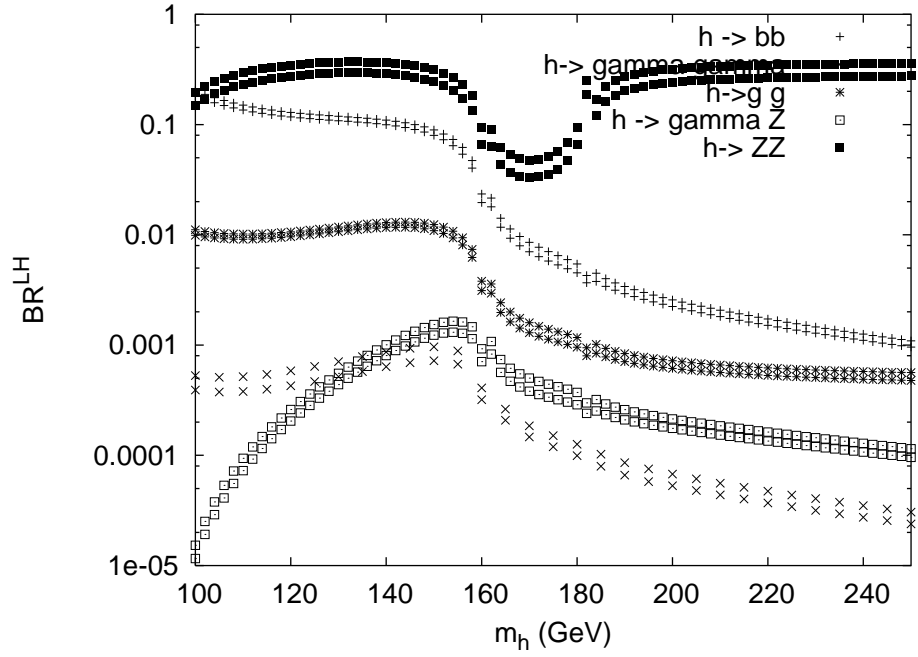}
\caption{ The branching fractions for a wide range of parameters, fixed  $f=1$ TeV versus
  the Higgs boson mass. }
\end{figure}

\end{document}